# Learning to predict test effectiveness

Morteza Zakeri-Nasrabadi[1]

[1] Ph.D. Student,
School of Computer Engineering, Iran University of Science and Technology, Tehran, Iran.
(*morteza_zakeri@comp.iust.ac.ir*)

Saeed Parsa[2],*

[2] Associate Professor,
School of Computer Engineering, Iran University of Science and Technology, Tehran, Iran.
(*parsa@iust.ac.ir*)

*Abstract*— The high cost of the test can be dramatically reduced, provided that the coverability as an inherent feature of the code under test is predictable. This article offers a machine learning model to predict the extent to which the test could cover a class in terms of a new metric called *Coverageability*. The prediction model consists of an ensemble of four regression models. The learning samples consist of feature vectors, where features are source code metrics computed for a class. The samples are labeled by the Coverageability values computed for their corresponding classes. We offer a mathematical model to evaluate test effectiveness in terms of size and coverage of the test suite generated automatically for each class. We extend the size of the feature space by introducing a new approach to defining sub-metrics in terms of existing source code metrics. Using feature importance analysis on the learned prediction models, we sort source code metrics in the order of their impact on the test effectiveness. As a result of which, we found the class strict cyclomatic complexity as the most influential source code metric. Our experiments with the prediction models on a large corpus of Java projects containing about 23,000 classes demonstrate the Mean Absolute Error (MAE) of 0.032, Mean Squared Error (MSE) of 0.004, and an $R^2$-score of 0.855. Compared with the state-of-the-art coverage prediction models, our models improve MAE, MSE, and an $R^2$-score by 5.78%, 2.84%, and 20.71%, respectively.

## 1 Introduction

Predicting test effectiveness and quality is vital because finding all bugs in a program is not a decidable problem in general [1]. Test effectiveness depends on the extent to which the selected test cases cover the code under test. Practically, determining the extent of the test coverage after each change to the code conducts programmers towards developing testable code. The point is that by assigning enough time and resources for test data generation, the generated test data could ultimately cover all the feasible paths of a program. However, measuring test coverage after each change to the code is costly and time-consuming because the coverage reveals only after running the code with the selected test cases [1]. In addition, the code has to be executable.

The extent to which test cases could cover a source code depends on the code quality attributes measured by source code metrics. However, empirical studies show that the correlation between code coverage and most software metrics is not strong enough [2]. Indeed, different test suites may provide the same coverage for a given source code. This article demonstrates a strong correlation between software metrics and the coverage attained by a single test case. To this aim, we introduce a new concept named Coverageability. Coverageability is the extent to which a source code is expected to be covered by any test suite, considering the test budget. This article presents a mathematical model to compute Coverageability in terms of the coverage attained when running the class under test. We build machine learning regressor models to map classes represented as vectors of metrics to their Coverageaiblity computed by the mathematical model. There are three main challenges while adopting machine learning to build a Coverageability prediction model.

First, the primary challenge with constructing the machine learning model is selecting metrics to build the feature space [2], [3]. Initially, 54 known software metrics were collected. The feature space was extended by adding statistical and lexical metrics to obtain better performance. In this way, the number of features increased from 71 to 296 metrics. The permutation importance technique [4] was applied to sort the metrics in order of their importance in the prediction of Coverageability.

The second challenge is to provide an adequate number of samples to train and test the machine learning model [5]. Each sample is a vector representation of a class where each element of the vector indicates a metric. The samples are labeled with the Covergeability of their corresponding class computed by the mathematical model. The mathematical model computed Coverageability using the size and the geometric average of statement, branch, and mutation coverage of the test suite generated by running EvoSuite on 110 open source projects in the SF110 corpus [6]. We created 23000 samples representing 23000 Java classes of the Java projects.

---

* Corresponding author

Finally, the third challenge is selecting the most appropriate combination of test adequacy criteria, representing test effectiveness [7]. It has been shown that combining coverage criteria provides a relatively more accurate indication of test effectiveness than using only a single criterion [8]. For instance, the EvoSuite test data generator tool [9] combines eight code coverage metrics when generating tests. Our experiments with EvoSuite suggest that combining statement, branch, and mutation coverage can boost test effectiveness. In summary, the primary contributions of this paper are as follows:

1. To introduce a novel software quality metric, Coverageability, considering test adequacy criteria and test budget to provide a proxy for test effectiveness measurement.
2. To offer a machine learning model, predicting Coverageability in terms of code metrics statically.
3. To introduce the concept of sub-metrics and lexical metrics as informative features that enhance the effectiveness of learning models in addition to traditional object-oriented (OO) metrics.
4. To automatically designate the top 15 metrics affecting Coverageability of the class under test.

Our empirical study indicates the Mean Absolute Error (MAE) of 0.032, Mean Squared Error (MSE) of 0.004, and an $R^2$-score of 0.855 evaluation error for our regression ensemble model. The proposed Coverageability models have improved the MAE by 5.78%, MSE by 2.84, and $R^2$-score by 20.71% compared with the state-of-the-art machine learning approach [7]. Our Coverageability dataset is publicly available on Zenodo [10]. Our results and replication package can be found on *https://m-zakeri.github.io/ADAFEST/*.

The rest of this article is organized as follows: In Section 2, related works on measuring source code testability are discussed. Section 3 introduces the Coverageability formulation and our measurement approach. Section 4 describes our experimental study, followed by answering relevant research questions. In Section 5, the possible threats to validity are discussed. Finally, in Section 6, the conclusion and future works are discussed.

## 2 Related work

Studies on software testability measurement are of two categories: The first category deal with measuring test effort in terms of the cost and budget. The second category concerns the measurement of test effectiveness. Accurate and efficient measurement of both categories would be precious for software engineers. This section reviews the current approaches to measure test effectiveness and shows promising results.

### 2.1 Test effectiveness measurement

Code coverage criteria assess the quality of a test data set as the percentage of code it covers. The higher the coverage, the higher the probability of test success. Therefore, test effectiveness relies on the extent of the test coverage. Frankl and Iakounenko [11] have shown that the likelihood of detecting a fault grows sharply as coverage increases. They concluded that code coverage is an appropriate means to determine test effectiveness. Cruz and Eler [12] have investigated the correlation between test coverage, mutation score, and object-oriented systems metrics. Their preliminary results show that some C&K metrics [13] strongly affects line coverage and mutation score. However, they have not performed further statistical analysis to demonstrate any result related to data distribution for prediction tasks. Despite the statement that software metrics do not capture some relevant factors involved in test case development [14], we believe that considering the test effort in terms of size of the test suite and test budget, can amplify the correlation between source code metrics and test effectiveness, which is required to create accurate prediction models.

Oliveira et al. [15] have measured test effectiveness in terms of the branch coverage achieved by the test data provided using search-based and random test data generation tools. They have trained a decision tree classifier to map source code metrics to the coverage achieved by the test suites generated by different algorithms for each class under test. Three significant features affecting the selection of test data generation techniques have been reported, including Response For a Class (RFC), the coupling between object classes (CBO), and the number of methods (NOM) in a class. However, they have solely considered branch coverage to measure test effectiveness.

Catolino et al. [16] have conducted an empirical study on the relationship between development teams' experience the number of assertions per line of code in a test class. They have concluded that test effectiveness is highly affected by the developer's experience. In contrast, we show that test effectiveness is mainly affected by metrics of the class under test.

Grano et al. [17] have interviewed software developers and found that available test metrics effectively characterize poor-quality tests while they are limited in distinguishing high-quality ones. They have concluded that some novel metrics and mechanisms are required to support developers with the assessment of test code quality. Grano et al. [18] have shown that source code quality indicators can estimate test-case effectiveness without executing the tests in terms of mutation score. They have used both the production code metrics and test code metrics to build a boolean classifier predicting "effective" or "non-effective" tests for a given class under test. Grano's approach works only in the presence of test code and rigidly classifies tests in one of the two mentioned categories. In contrast, our Coverageability prediction approach does not require any tests and estimate test effectiveness in a continuous and interpretable range.

Aghamohammadi et al. [19] have introduced a new code coverage criterion, called statement frequency coverage, to measure test suite effectiveness. They show that this metric outperforms statement and branch coverage. Statement frequency coverage counts the number of times each statement in the class under test is executed during the test. Unlike, Coverageability, statement frequency coverage requires software under test to be executed to assess the effectiveness of the test.

In the most recent work, Terragni et al. [2] have proposed an approach to measure testability considering both test effectiveness and test effort. They have computed test effort and test quality metrics for each test class, TC, and then normalized the test effort with the actual test quality of TC. Their normalization is based on the intuition that testing effort grows with an increased test quality. They have concluded that normalizing test effort with test quality increases the correlation between class-level source code metrics and test effort. However, they have not discussed how the result of this correlation must be interpreted. In this paper, we act inversely, *i.e.*, normalize the test effectiveness with test effort. The mathematical expectation of the statement and branch coverage is considered as a concrete proxy for test effectiveness. In addition, the size of the test suite is used to determine the required testing effort. The test coverage to test effort ratio is defined as Coverageability. More Coverageability means accessing more effective test suites with a lower test budget is possible for a given class.

## 2.2 Machine learning in software engineering tasks

Machine learning approaches are about creating mathematical models that learn from available data to make decisions, predictions, categorizations, and recommendations. In the last two decades, machine learning techniques have been applied to different aspects of software engineering and software testing [20], mainly test data generation [21], test oracle generation [22], fault prediction [23], and fault localization [24]. The typical approach for using machine learning to address a problem in software engineering is to extract some features from an artifact and relate these features to a desirable target. For example, most related works extract source code metrics in fault prediction and connect them to a fault proneness value [23].

Fontana and Zanoni [25] have used machine learning to classify the severity of code smells based on source code metrics. Aniche et al. [26] have investigated the effectiveness of machine learning algorithms in predicting software refactorings opportunities based on source code metrics. It seems that there is a high potential capacity for machine learning techniques to understand and solve difficult software engineering tasks automatically utilizing existing software artifacts such as source code. This article leverages the well-known supervised machine learning technique, regression learning, to learn to predict Coverageability as an actual and explainable measure quantifying software testability.

## 2.3 Machine learning for test effectiveness measurement

As one of the early studies on test effectiveness estimation with machine learning techniques, Daniel and Boshernitsan [5] have attempted to learn to predict code coverage provided by automated testing tools. They trained a decision tree classifier on 11 open-source Java projects to predict code coverage of the method under test in three automated testing tools: Agitator [27], Agitator with Mockitator engine [28], and Randoop [29]. However, their model could only predict the code coverage as discrete high (100% coverage) and low (0% coverage) categories, which is not very interesting.

Grano et al. [7] have applied regression models to predict branch coverage of Class Under Test (CUT) on seven open-source Java projects by two unit testing tools, EvoSuite [30] and Randoop [29]. Despite the ability to predict the numerical value of code coverage, the performance of their proposed models, precisely the mean absolute error (MAE) and $R^2$-score are relatively low. Their proposed approach mainly suffers from the problems discussed in Section 1. One important problem is that branch coverage alone is not a good indicator for code coverage [31]. We empirically observed a weak correlation between achievable branch coverage and source code metrics resulting in poor learning and prediction performance. On the other hand, Salahirad et al. [32] have shown that branch coverage of a test suite is the factor that is most strongly tied to fault detection. Therefore, it is crucial to consider code coverage value while measuring software testability.

EvoSuite is a state-of-the-art tool for testing Java classes [30]. It provides many test adequacy criteria using evolutionary search-based algorithms. Using automatic test data generation, we ensure that different Java classes are tested under identical conditions. To the best of our knowledge, this is the first study in testability measurement, which controls parameters such as fixing the test budget to measure the test effectiveness fairly.

# 3 Proposed method

This section formally defines the Coverageability of a class under test (CUT) and explains a Coverageability measurement framework used to estimate the Coverageability based on source code metrics.

## 3.1 Coverageability mathematical model

Coverageability is an inherent feature of source code. It is a measure of test effectiveness. We define Coverageability, $C_\mu$, of a class, $X$, as the extent to which a test suite, $\tau(X, C)$, can satisfy test requirements considering the test budget, $B$. A test requirement is defined as a specific element of a software artifact that test data must cover or satisfy [1]. Given a set of test requirements, TR, and a test suite $\tau(X, C)$, the coverage level is the ratio of the number of test requirements satisfied by $\tau(X, C)$ to the size of TR. The coverage level, $C^{level}(X)$, could best determine the test effectiveness [1]. Coverageability, $C_\mu(X)$, of a class, $X$, is directly proportional to its coverage level:

$$C_\mu(X) \propto C_{TR}^{level}(X) \qquad (1)$$

According to Google [33] and IBM [34], statement and branch coverage are the most practical test adequacy criteria. However, due to the possible imbalance of statements among the branches, neither branch nor statement coverage can be a good indicator of test effectiveness on their own. For instance, ten branches in a source code may cover fewer statements than a single branch. Therefore, the statement and branch coverage criteria should complement each other when computing test effectiveness.

On the other hand, the Coverageability, $C_\mu(X)$, of a class X decreases as the number of test data, $|\tau(X,C)|$, required to achieve the coverage, C, increases:

$$C_\mu(X) \propto \frac{1}{|\tau(X,C)|} \qquad (2)$$

To be specific, it can be said that the Coverageability of class A is higher than that of B if fewer test data are required for A than for B to achieve a specific coverage. Generally, the size of the test suite is an influential factor in test effectiveness.

The mathematical expectation of the code coverage, $E(C^{sat}(X))$, is defined as the arithmetic mean of the statement coverage and branch coverage to overcome the insufficiency of code coverage criteria when used in isolation:

$$E(C^{sat}(X)) = \frac{\left(C_{ST}^{level}(X) + C_{BR}^{level}(X)\right)}{2} \qquad (3)$$

Finally, Coverageability is given by Equation 4:

$$C_\mu(X) = \frac{E(C^{level}(X)) \times b}{|\tau(X,C)|} \qquad (4)$$

where $1 \leq b \leq |\tau(X,C)|$ indicates the minimum number of influential test cases required to achieve an effective test, considering the test budget, $B$. By an influential test case, we mean a test case that increases the test suite coverage.

Equation 4 quantifies the value of Coverageability in the continuous interval of (0, 1]. Considering Equation 4, a class X the Coverageablity, $C_\mu(X)$, of class X reaches its maximum value, 1, when the size of the test suite to satisfy the branch coverage criterion is b.

The Coverageability of a module, $M$, can be defined as the arithmetic average of Coverageability of all classes within $M$:

$$C_\mu^M(M) = \frac{1}{n} \sum_{i=1}^{n} C_\mu(X_i) \qquad (5)$$

## 3.2 Coverageability measurement framework

The aim is to provide an ensembled regressor model to statically predict Coverageability at a class level and in terms of software metrics evaluated for the class. Our Coverageability measurement framework aims to learn a real-valued function $f$ that maps the vector of metrics $V$, computed for a class $X$ to the Coverageability value $C(X)$ with minimum possible error, *i.e.*, $f: V \in \mathbb{R}^n \rightarrow T \in \mathbb{R}$. The underlying assumption is that the Coverageability of a class, $X$, depends on its structural features, determined by the metrics.

Figure 1 illustrates the Coverageability measurement framework. It consists of two main phases of construction and application. The construction phase concerns the model construction, and the application phase deals with applying the model in practice. The construction and application phases, respectively, comprise 5 and 3 steps as follows:

1. **Collect source code repositories:** An extensive set of source codes is processed to create the required dataset. We used the SF110 corpus [16], consisting of 110 Java projects with more than 23,000 Java classes, to construct our Coverageability prediction model.

2. **Compute test criteria:** We used EvoSuite [6] to automatically generate test suites for 23000 Java classes in our source code repository. EvoSuite runs each class with its corresponding test suite to compute test requirements, including the branch and statement coverage. We then apply Equation 4 to compute the Coverageability of each class. It is worth mentioning that automatic test data generation is preferred; otherwise, we should consider human expertise as a factor affecting Coverageability.

3. **Compute static metrics:** A set of 256 metrics are used to quantify each Java class quality attribute. The vector of source code metrics computed for each class is labeled with the class Coverageability, computed by Equation 4. Each labeled vector is used as a sample for training and testing the Coverageability prediction model.

4. **Preprocessing:** The value of source code metrics falls into various ranges that are not suitable for most machine learning algorithms. Besides, some classes may only consist of data fields whose Coverageability is inherently high. During the preprocessing, the value of metrics is scaled into a specific range by applying standardization techniques, and data classes are removed from the dataset.

5. **Training:** Well-known machine learning algorithms are examined to train prediction models for Coverageability, and the best model, *i.e.*, the model with a minor error, is chosen.

Steps 6 and 7 of the online phases are the same as steps 3 and 4, respectively. These steps apply to any new Java class to estimate its Coverageability. Step 8 invokes the learned model to predict the Coverageability based on preprocessed metrics.

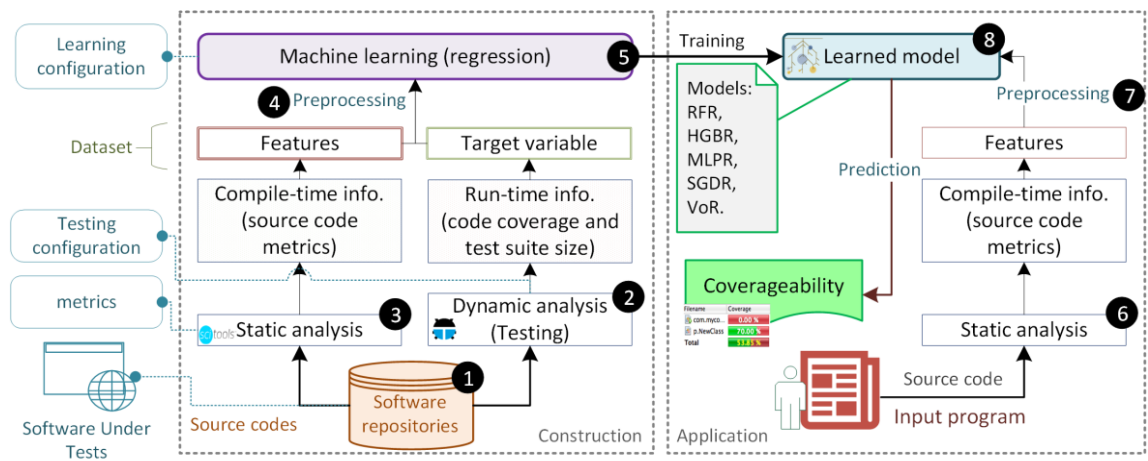

Figure 1. Coverageability measurement framework

## 3.3 Metrics extraction

An extensive set of source code metrics is extracted and computed for each Java class. Metrics constitute the feature space to construct and use our machine learning model. We propose the two categories of sub-metrics and lexical metrics to enhance the feature space effectively.

### 3.3.1 Object-oriented metrics

Numerous object-oriented metrics are proposed in the literature to quantify different source code characteristics or quality attributes. At the same time, only a few of them have been utilized to measure test effectiveness. We compute all well-known object-oriented metrics, including C&K metrics [13], [35], HS metrics [36], MOOD metrics [37], QMOOD metrics [38], and MTMOOD metrics [39]. We also consider Custom Metrics introduced in [40]. A complete list of object-oriented metrics is given in Table 1.

Table 1 classifies software metrics by their quality attribute at the two levels of class and package components. The quality attributes are size, complexity, cohesion & coupling, visibility, and inheritance. The core elements of object-oriented programs are methods, reference types (including classes, interfaces, and enums), source files, and packages or namespaces. Since Coverageability is defined at the class level, we use metrics defined for the other elements to determine class attributes.

Table 1. Selected source code metrics with different granularities and subjects.

| Entity | Quality attribute | | | | | |
|---|---|---|---|---|---|---|
| | Size | Complexity | Cohesion and coupling | Visibility | Inheritance | Sum |
| Class (CS) | **CSLOC** (36) **CSNOST** (36) CSNOSM CSNOSA CSNOIM CSNOIA CSNOM CSNOMNAMM CSNOCON **CSNOP** (10) | **CSCC** (48) **CSNESTING** (4) **CSPATH** (10) **CSKNOTS** (10) | LOCM CBO RFC FANIN FANOUT DEPENDS DEPENDSBY ATFD CFNAMM DAC NOMCALL | CSNODM CSNOPM CSNOPRM CSNOPLM CSNOAMM | DIT NOC NOP NIM NMO NOII | 36 (183) |
| Package (PK) | **PKLOC** (15) **PKNOST** (15) PKNOSM PKNOSA PKNOIM PKNOIA PKNOMNAMM PKNOCS PKNOFL | **PKCC** (48) **PKNESTING** (4) | — | PKNODM PKNOPM PKNOPRM PKNOPLM PKNOAMM | PKNOI PKNOAC | 18 (96) |
| Sum | 19 (126) | 6 (124) | 11 | 10 | 8 | 54 (279) |

### 3.3.2 Sub-metrics

This research improves the predictability of the conventional penalized regression models by effectively enhancing the feature space with statistical sub-metrics and lexical metrics and also using the feature generation techniques [41]. Sub-metric is a new concept introduced in this article. Sub-metrics are derived from a metric by applying statistical operators, including sum, mean (AVG), minimum (MIN), maximum (MAX), logarithm (LOG), and standard deviation (SD) to the metric. We apply statistical operators to the nine metrics, bolded in Table 1.

All the metrics defined for an Object-oriented element can be computed for the parent of that element by applying the statistical operations. For instance, a class cyclomatic-complexity (WMC) is computed as the sum of its methods Cyclomatic Complexity (CC). Some methods of a class can be ignored before applying the statistical operation. For instance, we ignore the accessor and mutator methods when computing a class CC as the sum of its methods CC. Besides, some metrics, such as CC, have several definitions that result in different values. For instance, Cyclomatic Complexity has four different definitions: CC, CC-strict, CC-modified, and CC-essential. Constructing sub-metrics provides the ability to analyze and study the impact of each primary metric in more detail. For example, we can answer whether a class with a high CC variance is relatively more testable.

Figure 2 summarizes our proposed method to consider sub-metrics for a given metric, such as CC. It illustrates a web of the sub-metrics construction for CC at the class level. As shown in Figure 2, we could define $6 \times 2 \times 4 = 48$ different effective sub-metrics to quantify a class CC.

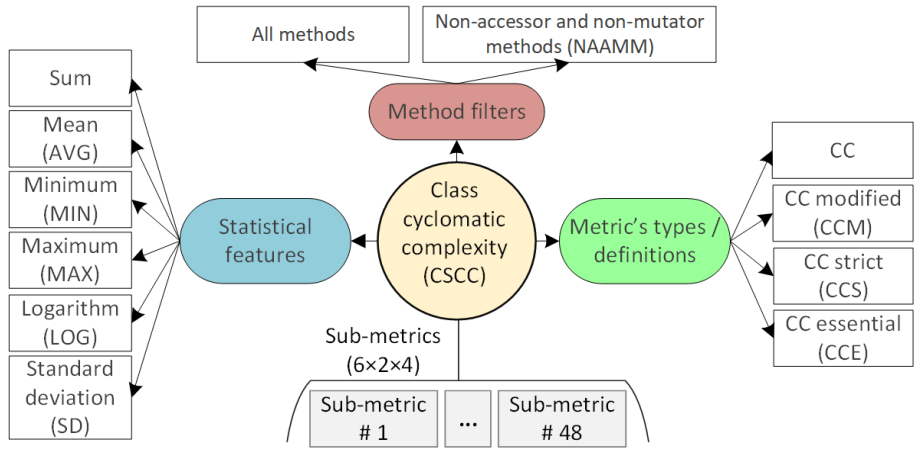

Figure 2. Web of sub-metrics for the class Cyclomatic Complexity

### 3.3.3 Lexical metrics

We introduce a new set of metrics called lexical metrics to capture lexical properties, such as the number of variables and operators used in a class. In this way, we could study the impact of fine-grained lexical properties on the accuracy of our regression models. Table 2 provides a list of 17 lexical metrics and their relevant quality attributes. Lexical metrics are defined and computed for each file in a program to consider the impact of a class's lexical features on its Coverageability, *e.g.*, the number of tokens. In total, we used 71 (54+17) metrics and extended 11 of them with sub-metrics, which resulted in 296 (279+17) different metrics.

Table 2. Lexical metrics and their definitions

| Metric's abbreviation | Metric name | Relevant quality attribute |
| --- | --- | --- |
| NOTK | Number of tokens | Size |
| NOTKU | Number of unique tokens | Size |
| NOID | Number of identifiers | Size |
| NOIDU | Number of unique identifiers | Size |
| NOKW | Number of keywords | Size |
| NOKWU | Number of unique keywords | Size |
| NOASS | Number of assignments | Complexity |
| NOOP | Number of operators without assignments | Complexity |
| NOOPU | Number of unique operators | Complexity |
| NOSC | Number of semicolons | Size |
| NODOT | Number of dots | Complexity |
| NOREPR | Number of return and print statements | Size |
| NOCJST | Number of conditional jumps | Complexity |
| NOCUJST | Number of unconditional jumps | Complexity |
| NOEXST | Number of exceptions | Complexity |
| NONEW | Number of *new* objects instantiation | Coupling |
| NOSUPER | Number of *super* calls | Inheritance |
| Sum | 17 | |

## 3.4 Dataset preparation

The dataset preparation consists of two steps: arranging source code metrics as feature vectors and then preprocessing the vectors in a suitable shape to construct the Coverageability prediction model.

### 3.4.1 Data representation

The feature space comprises 296-dimensional vectors of 296 elements, each element representing the value of a sub-metric/ metric computed for a class. Each vector is labeled with the Coverageability value, computed for its corresponding class. Equation 4 computes a class Coverageability. For each class, $X$, in a given project, feature vector, $V_X$, is defined as:

$$V_X = <M_{P(X)}, M_{F(X)}, M_X> \qquad (6)$$

where $M_{P(X)}$, $M_{F(X)}$, and $M_X$ Indicate package-level metrics, file-level lexical metrics, and class-level metrics, respectively. Class level sub-metrics are computed based on method level metrics, and package level sub-metrics are computed based on the class metrics. Apparently, the package level metric, $M_{P(X)}$, is the same for all the classes, $X$, in a package. Lexical metrics are considered at the file level. Therefore, if there is more than one class, $X$, in a given File, $F$, the file level metrics, $M_{F(X)}$, for all the classes is the same. We use 17 file-level metrics listed in Table 2.

Supervised learning requires each feature vector to have at least one variable as a target for training. Therefore, for each class, $X$, a target vector, $T_X$, consists of the four targets addressed in Equations 3 and 4. The vector is as follows:

$$T_X = <C_{ST}^{level}(X), C_{BR}^{level}(X), E(C^{level}(X)), C_\mu(X)> \qquad (7)$$

Machine learning algorithms are applied to build a prediction model for each target separately. The prediction results evaluate the correctness of the learned model empirically.

### 3.4.2 Data preprocessing

Figure 3 illustrates the data preprocessing steps. First, we removed samples considered outliers when applying the LOF algorithm [42]. We also removed Simple classes (classes with $LOC < 5$) and Data classes, *i.e.*, classes containing only data fields, mutators, and accessor methods. Our primary observations show that simple and data classes are inherently Coverageable. A few test data cover most of the SF110 simple and data classes.

After data cleaning, the dataset splits into two sets of train and test samples. Then feature standardization and feature selection are applied to the train and test datasets consecutively. Some machine learning algorithms, such as neural networks (MPLs), are sensitive to the feature value ranges. Standardization scales metrics as features into a standard range. We applied a Robust Scaling standardization method. This method ignores the feature median and scales its values in different samples based on their Interquartile Quantile Range (IQR), *i.e.*, the range between the 1st and the 3rd quartile. Centering and scaling occurred independently on each feature by computing the relevant statistics for all samples in the training set.

We compute 296 metrics, which result in a high-dimensional feature space. These metrics may correlate together or even be uncorrelated with the Coverageability. Feature selection algorithms are applied to eliminate redundant metrics and boost learning performance by selecting the most informative metrics. The univariate feature selection is applied to select the best features based on univariate statistical tests [43]. Our experiments are performed with and without feature selection to discover the impact of automatic metric selection on Coverageability prediction.

It is worth noting that another approach to reducing dimensionality is feature extraction, in which the new features are created based on the current features that are more informative and have less correlation together. The most common feature extraction technique is Principal Component Analysis (PCA) which transforms the data into a new coordinate system such that the variance of each dimension maximizes. We did not apply feature extraction since it introduces new synthetic features that are not interpretable as software metrics.

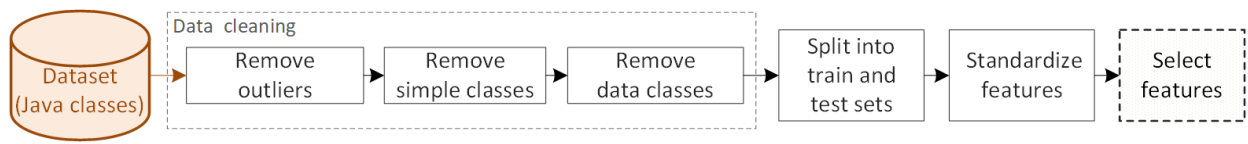

Figure 3. Preprocessing steps pipeline.

## 3.5 Learning to predict Coverageability

Regression learning techniques are utilized to predict Coverageability. Machine learning regression techniques learn to predict a continuous value as output. We explain our learning and inference algorithms, their evaluation, and their implementation.

### 3.5.1 Learning algorithms

The SGDR, MLPR, RFR, and HGBR algorithms are used to predict Coverageability. We chose them because they are off-the-shelf machine learning algorithms that belong to different families and have been of great use to many real-world problems [7], [4]. Below is a brief description of each of these models:

1. **Stochastic Gradient Descent Regressor (SGDR)** [44]: It is a linear regressor with convex loss functions such as least-squares and is well suited for regression problems with many training samples. SGDR fits a linear model on the given data. SGDR provides relatively more accurate results, provided that there is a linear relationship between the source code metrics and the software quality attributes. The other learning algorithms are non-linear regression.

2. **Multi-Layer Perceptron Regressor (MLPR)** [45]: It is a class of feed-forward neural networks with a deep architecture that consists of multiple layers of many computational neurons interconnected in a directed acyclic graph (DAG). MLP builds a function for mapping input to output and works well in most regression tasks. The weight of each connection in DAG is learned during the training.

3. **Random Forest Regressor (RFR)** [4]: It is an ensemble learning method that constructs many decision trees on bootstrapped data. Each tree is built on a subset of original data and features, resulting in a more robust model than the Decision Tree. An important hyperparameter that affects the quality of learning is the number of decision trees used as weak learners.

4. **Histogram-based Gradient Boosting Regressor (HGBR)** [46], [47]: It is a kind of Gradient Tree Boosting that uses decision tree regressors as weak learners while trying to overcome the significant problem of typical gradient boosting methods, *i.e.*, the efficiency of training on a large dataset and many numbers of weak learners. In the Histogram-based Gradient Boosting technique, the process of training the trees which are added to the ensemble is dramatically accelerated by discretizing (binning) the continuous input variables to a few hundred unique values. Each decision tree then operates upon the ordinal bucket instead of specific values in the training dataset.

We also built a voter regressor (VoR) on top of the other four models to combine their outcomes. The outcomes are combined with different weights to detect the best result on the training set. Figure 4 depicts the voting process. The final result is the weighted average of the predictions provided by the trained model predictions.

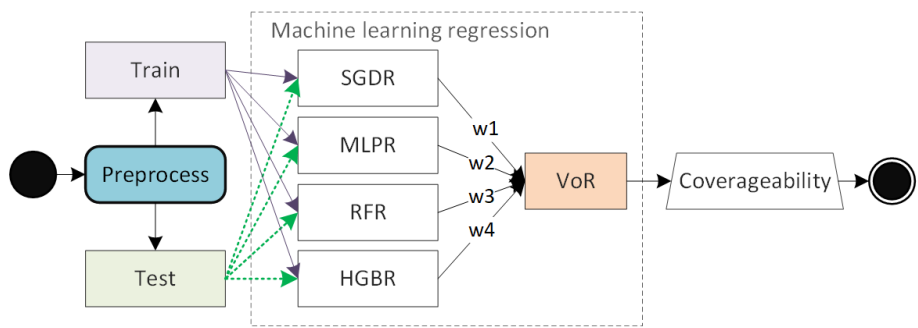

Figure 4. Voting process

### 3.5.2 Learning process

The learning algorithms described in Section 3.5.1 require different hyper-parameters, *i.e.*, parameters that are not learned and are determined before the training. To this aim, a *grid search* strategy with *cross-validation* [48] is employed to find the optimal hyperparameters to train each of the models. The grid search strategy exhaustively goes through manually specified subspaces of the learning algorithm hyperparameter space and selects the best configuration based on a given performance metric.

Table 3 shows the most critical hyperparameters for each model and their candidate values to be optimized during the grid search process. We selected the literature candidate values, considering the experts' recommendations and best practices [49]. In the case of the VoR model, the only available hyperparameter is the weight that is used to weigh the prediction results provided by the four Coverageability prediction models before averaging the results.

Table 3. Hyperparameters of each model to be used in the grid search

| Model | Hyper-parameter name in Scikit-learn | Searching values as Python statements |
|---|---|---|
| SGDR | • loss<br>• penalty<br>• learning_rate<br>• max_iter | `['squared_loss', 'huber']`<br>`['l2', 'l1', 'elasticnet']`<br>`['invscaling', 'optimal', 'constant', 'adaptive']`<br>`range(50, 500, 50)` |
| MLP | • hidden_layer_sizes<br>• activation<br>• learning-rate<br>• epochs | `[(128, 64), (256, 100), (512, 256, 100)]`<br>`['relu', 'tanh', 'logistic',]`<br>`['constant', 'adaptive']`<br>`range(100, 500, 50)` |
| RFR | • n_estimators<br>• criterion<br>• max_depth<br>• min_samples_split | `range(50, 500, 50)`<br>`['mse', 'mae']`<br>`range(3, 50, 1)`<br>`range(2, 30, 2)` |
| HGBR | • loss<br>• max_depth<br>• min_samples_leaf<br>• max_iter | `['least_squares', 'least_absolute_deviation']`<br>`range(3, 50, 1)`<br>`range(5, 50, 10)`<br>`range(50, 500, 50)` |
| VoR | • weights (w1-w4 respectively) | `[None, [0, 1/3, 1/3, 1/3], [0, 1/6, 2/6, 3/6]]` |

### 3.5.3 Inference algorithm

The Coverageability of a new class is predicted by extracting its metrics and fed into a learned regression model. Algorithm 1 estimates the Coverageability of class, $X$, by accepting the class source code, $SC$, one of the four learned models, $M$, the distribution, $D$, and the parameters, $\theta$.

The algorithm first checks whether $X$ is a Simple or Data class, using simple rules with predefined thresholds. If $X$ is a Simple or Data class, Coverageability will be equal to one; otherwise, the model is asked to estimate Coverageability. If the predicted value is out of range [0, 1], an exception will be raised since regression models may predict any value within an arbitrary range, specifically when their input (independent variables) is out of the learned distribution.

The main goal of predicting Coverageability with Algorithm 1 is to check whether the code is ready for the test. If the model predicts low Coverageability (*e.g.*, < 0.50), testing must be postponed until the SUT testability is improved. In other words, measuring Coverageability reduces the time and cost of *inappropriate testing*.

---

***Algorithm 1.** CalculateClassCoverageability*

**Inputs:** *ProjectSourceCode SC, Class X, LearnedModel M(D; θ)*

**Output:** *float Coverageability*

```
metrics = ComputeMetric(SC, X);   /* Extract all metrics for the class X */
a = metrics["CSLOC"]              /* Let a = lines of code (CSLOC) of X */
b = metrics["CSNOMNAMM"]          /* Let b = no. methods appart from the Accessors and Mutators
                                            Methods (CSNOMNAMM) of X */
c = metrics["CSNOIA"]             /* Let C = no. instance attributes (CSNOIA) of X */
d = metrics["CSNOSA"]             /* Let d = no. static attributes (CSNOSA) of X */

if a < 5:   /* The rule to check whether X is a Simple class */
    Coverageability = 1
else if b == 0 & (c + d) > 0:   /* The rule to Check whether X is a Data class */
    Coverageability = 1
else:
    Coverageability = M.predict(metrics, D; θ)   /* Predict Coverageability of the non-trivial class X */
if Coverageability >1.0 or Coverageability < 0:
    raise "Prediction Error"
return Coverageability
```

---

### 3.5.4 Evaluation metrics

Regression models are mainly evaluated based on the error metrics, measuring the difference between the actual value and the predicted value. For each model, we measure and report four typical error metrics, including Mean Absolute Error (MAE), Mean Squared Error (MSE), Root Mean Squared Error (RMSE), Mean Squared Logarithmic Error (MSLgE), and Median Absolute Error (MdAE). These metrics are given by Equation 8 – 12. In all equations, $\hat{y}_i$ is the predicted value of the $i^{\text{th}}$ sample, $y_i$ is the corresponding actual value for $\hat{y}_i$ and $n$ is the number of samples. A less error value means a better model.

$$\text{MAE}(y, \hat{y}) = \frac{1}{n}\sum_{i=0}^{n-1}|y_i - \hat{y}_i| \qquad (8)$$

$$\text{MSE}(y, \hat{y}) = \frac{1}{n}\sum_{i=1}^{n-1}(y_i - \hat{y}_i)^2 \qquad (9)$$

$$\text{RMSE}((y, \hat{y}) = \sqrt{\text{MSE}((y, \hat{y})} \qquad (10)$$

$$\text{MSLgE}(y, \hat{y}) = \frac{1}{n}\sum_{i=0}^{n-1}(\log_e(1 + y_i) - \log_e(1 + \hat{y}_i))^2 \qquad (11)$$

$$\text{MdAE}(y, \hat{y}) = \text{median}(|y_1 - \widehat{y_1}|, ..., |y_n - \widehat{y_n}|) \qquad (12)$$

Another performance metric is the coefficient of determination score or $R^2$-score, given by Equation 13. The proportion of the variance in the dependent variable is predictable from the independent variable(s). A constant model that always predicts the mathematical expectation of y, denoted by $\bar{y}$, disregarding the input features, would get an $R^2$-score of 0.0. A higher $R^2$-score means better model performance.

$$R^2(y, \hat{y}) = 1 - \frac{\sum_{i=1}^{n}(y_i - \hat{y}_i)^2}{\sum_{i=1}^{n}(y_i - \bar{y})^2} \quad (13)$$

### 3.5.5 Implementation

The entire Coverageability measurement framework is in Python, which leads to easy debugging and accessibility. Preprocessing and machine learning algorithms are implemented using Scikit-learn [49], a free python data analysis library. The framework produces all required material for any Java project by receiving its source code at the input. To this aim, source code metrics are extracted using Understand API for Python [50]. Understand analyzes a given project and creates a database file containing the symbol table and other program dependencies. It also provides Python APIs to query the database. Therefore, A python module is developed upon Understand API to compute all metrics in Table 1 and Table 2. Sub-metrics are built on their base metrics using statistical functions and suitable programs' element filtering. To process an extensive collection of Java projects, we utilized the command-line scripts to analyze and create all database files, then load each database and extract metrics for each class in the database. Figure 5 shows the component diagram of the implemented tool.

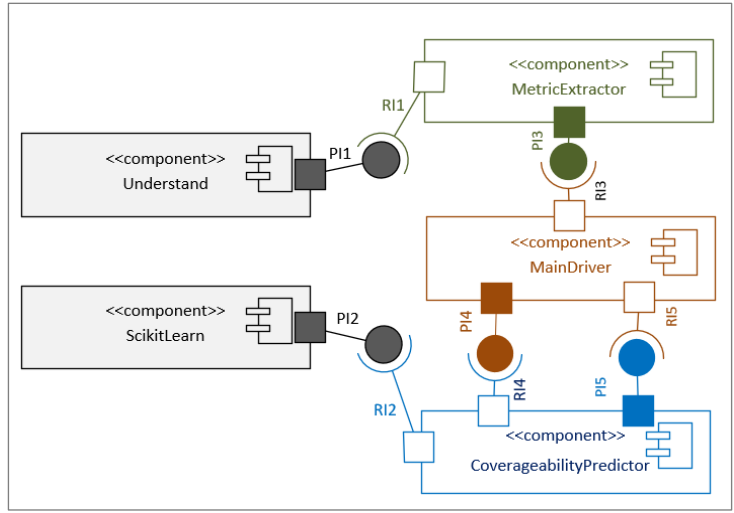

Figure 5. Component diagram of Coverageability measurement framework

## 4 Experiments

The main goal of our experiments is to discover how well Coverageability can be predicted based on source code metrics using the proposed measurement framework for real-world software projects. Besides, we aim to identify the most crucial source code metrics affecting Coverageability.

### 4.1 Research questions

The following research questions are defined to conduct our experiment in achieving these goals:

**RQ1:** What is the performance of learning models in predicting Coverageability? Which model has the best performance?

**RQ2:** What are the impacts of sub-metrics, file-level metrics (lexical metrics), and package-level metrics on predicting class Coverageability?

**RQ3:** What is the advantage of measuring Coverageability compared to predicting test adequacy criteria such as statement and branch coverage?

**RQ4:** What are the most influential source code metrics affecting Coverageability? To what extent do they affect Coverageability?

Concerning RQ1, we conduct a comparative analysis of the performance of five learning algorithms discussed in Section 3.5.1 to find the best machine learning model. Five datasets, listed in Table 4, are created with different metrics to answer RQ2.

In response to RQ3, we decided to repeat the Coverageability learning process on all datasets for each test adequacy criterion, including the statement and branch coverage. We then compared the result of predictions with Coverageability prediction to understand the usefulness of Coverageability in practice. Since Coverageability uses the combination of statement and branch coverage while considering the size of the test suite, source code metrics must predict it more accurately.

Finally, to answer RQ4, the well-known model inspection technique, permutation importance is applied to the best Coverageability prediction model. Permutation importance shuffles the values of a single feature and then asks a learned model to make predictions employing the resultant dataset with the shuffled column. Using these predictions and the actual target values determine how much the model suffered from shuffling. That performance deterioration denotes the importance of the shuffled feature.

Table 4. Dataset used in experiments

| Dataset | Description | Number of metrics |
|---|---|---|
| DS1 | The primary dataset includes all metrics | 296 |
| DS2 | DS1 + automatic feature selection | 15 |
| DS3 | DS1 + package-level metric elimination | 194 |
| DS4 | DS3 + file-level (lexical metrics) elimination | 177 |
| DS5 | DS1 + all sub-metrics elimination | 71 |

## 4.2 Experimental setup

All experiments were performed on a Windows 10 (x64) operating system with an Intel® Core™ i7 6700HQ CPU and 16GB of RAM. All machine learning models are trained and tested on all datasets after applying preprocessing steps discussed in Section 3.4.2. Constant b, the minimum number of test cases according to test budget, in Equation 4, set to 1, before computing the actual Coverageability value, determined by running EvoSuite [30].

SF110 [6] contains 110 Java projects with more than 23,000 classes, appropriated for machine learning tasks. A grid search with five-fold cross-validation is applied to select the best configuration of the hyperparameters. 75% of samples in each dataset are devoted to training, and the remaining samples are used for testing. The same ratio is used to divide the training set into train and validation sets for cross-validation. We apply the shuffle split technique for cross-validation. The technique does not guarantee that all folds are different, and it is the right choice for a large dataset. The shuffle split technique samples the entire dataset randomly during each iteration of the five-fold cross-validation to generate train and test sets.

The RMSE, given in Equation 10, has been used to score and rank the models during the cross-validation process. The model with the lowest RMSE is selected to be applied to the test set. A check-point mechanism was applied to save the best model found during the gird search. The entire training set was then used to retrain a final model with the obtained parameters and evaluate it on the test set.

## 4.3 Model and dataset analysis

This section describes the grid search and cross-validations applied to each of the four datasets in Table 4 to find the best prediction model and set of metrics to predict Coverageability.

### 4.3.1 Hyperparameter tuning

Table 5 shows the result of hyperparameter tunning with grid search on each Coverageability dataset. The best value of each hyperparameter between the values in Table 3 is shown for each dataset. Regarding the weight of SGDR being zero in the last row of Table 5, there is no linear relation between the metrics and Coverageability. Conversely, HGBR with a weight of 3.6 is the most influential model. Also, it is observed that the selected hyperparameters are the same for different data sets, which indicates that the hyperparameter models are not affected by the feature space.

Table 5. Models' configurable parameters and results of hyperparameter tunning

| Model | Best values on DS1 | Best values on DS2 | Best values on DS3 | Best values on DS4 | Best values on DS5 |
|---|---|---|---|---|---|
| SGDR | 'huber' | 'squared_loss' | 'huber' | 'huber' | 'huber' |
|  | 'l2' | 'l2' | 'l2' | 'l2' | 'l1' |
|  | 'invscaling' | 'adaptive' | 'invscaling' | 'invscaling' | 'invscaling' |
|  | 50 | 100 | 50 | 50 | 50 |
| MLP | (512, 256, 100) | (512, 256, 100) | (512, 256, 100) | (512, 256, 100) | (512, 256, 100) |
|  | 'tanh' | 'tanh' | 'tanh' | 'tanh' | 'tanh' |
|  | 'constant' | 'constant' | 'constant' | 'constant' | 'constant' |
|  | 100 | 50 | 100 | 50 | 50 |
| RFR | 100 | 100 | 100 | 100 | 100 |
|  | 'mse' | 'mse' | 'mse' | 'mse' | 'mse' |
|  | 30 | 20 | 20 | 20 | 40 |
|  | 2 | 2 | 2 | 2 | 2 |
| HGBR | 'least_squares' | 'least_squares' | 'least_squares' | 'least_squares' | 'least_squares' |
|  | 10 | 10 | 30 | 20 | 20 |
|  | 25 | 5 | 35 | 35 | 15 |
|  | 400 | 400 | 500 | 500 | 500 |
| VoR | [0, 1/6, 2/6, 3/6] | [0, 1/6, 2/6, 3/6] | [0, 1/6, 2/6, 3/6] | [0, 1/6, 2/6, 3/6] | [0, 1/6, 2/6, 3/6] |

### 4.3.2 Model performance

Table 6 shows the result of evaluating Coverageability prediction models with the best set of hyperparameters on each dataset using evaluation metrics discussed in Section 3.5.4. The best result for each model evaluation metric is marked with the *italic* font, and the best result of all models is bolded. If a learned model predicts a negative value, MSLgE, given in Equation 11, is undefined since the log function is not defined for negative numbers. The NaN symbol denotes the undefined value for MSLgE. The following result is observed:

- For the SGDR model, we found it cannot be trained on any of the Coverageability datasets. It means that there is no linear relationship between source code metrics and Coverageability.
- For MLPR and RFR models, DS3 denotes the best results for most evaluation metrics. The RFR regression absolutely performs better than the MLP model. RFR also never predicts a negative value for Coverageability. It shows that tree-based learning models are more suitable to use for Coverageability prediction than neural network models.
- The HGBR model on DS1 performed better than other datasets. However, its performance on DS3 is also very close to DS1, which indicates that DS3 contains a more suitable subset of metrics.
- Finally, the VoR model on DS3 has the best prediction performance among all models and datasets. It means that the combination of different non-linear learning models can still boost the final prediction result. The best result on DS3 shows that the package-level metrics are not suitable to predict class Coverageability; however, other metrics have positive impacts on predicting Coverageability.

Table 6. Performance of Coverageability prediction models on each dataset

| Model | Dataset | MAE | MSE | RMSE | MSLgE | MdAE | $R^2$-score |
|---|---|---|---|---|---|---|---|
| SGDR | DS1 | >>100 | >>100 | >>100 | NaN | 376.99 | <0 |
|  | DS2 | 0.0785 | 0.0148 | 0.1215 | NaN | 0.0548 | 0.5397 |
|  | DS3 | >>100 | >>100 | >>100 | NaN | 387.76 | <0 |
|  | DS4 | >>100 | >>100 | >>100 | NaN | 387.20 | <0 |
|  | DS5 | >>100 | >>100 | >>100 | NaN | 9.5474 | <0 |
| MLPR | DS1 | 0.0643 | 0.0098 | 0.0990 | NaN | 0.0430 | 0.6945 |
|  | DS2 | 0.0516 | 0.0093 | 0.0963 | NaN | *0.0249* | 0.7113 |
|  | DS3 | 0.0491 | *0.0075* | *0.0864* | NaN | 0.0258 | *0.7673* |
|  | DS4 | 0.0579 | 0.0086 | 0.0929 | NaN | 0.0361 | 0.7313 |
|  | DS5 | *0.0490* | 0.0078 | 0.0885 | NaN | *0.0249* | 0.7559 |
| RFR | DS1 | 0.0330 | 0.0054 | 0.0732 | 0.0027 | 0.0107 | 0.8329 |
|  | DS2 | 0.0383 | 0.0070 | 0.0835 | 0.0035 | 0.0122 | 0.7828 |
|  | DS3 | *0.0322* | *0.0050* | *0.0708* | **0.0026** | 0.0105 | *0.8440* |
|  | DS4 | 0.0332 | 0.0055 | 0.0740 | 0.0028 | 0.0108 | 0.8292 |
|  | DS5 | 0.0325 | 0.0053 | 0.0726 | 0.0027 | **0.0103** | 0.8357 |
| HGBR | DS1 | 0.0330 | *0.0049* | 0.0699 | *0.0025* | 0.0118 | *0.8478* |
|  | DS2 | 0.0410 | 0.0074 | 0.0862 | NaN | 0.0142 | 0.7685 |
|  | DS3 | *0.0325* | 0.0049 | *0.0699* | NaN | 0.0117 | 0.8477 |
|  | DS4 | 0.0341 | *0.0054* | 0.0733 | NaN | *0.0113* | 0.8326 |
|  | DS5 | 0.0327 | 0.0049 | 0.0702 | NaN | 0.0115 | 0.8465 |
| VoR | DS1 | 0.0331 | 0.0049 | 0.0697 | NaN | 0.0133 | 0.8485 |
|  | DS2 | 0.0402 | 0.0071 | 0.0840 | 0.0035 | 0.0142 | 0.7801 |
|  | DS3 | **0.0318** | **0.0047** | **0.0683** | NaN | 0.0119 | **0.8545** |
|  | DS4 | 0.0342 | 0.0052 | 0.0720 | NaN | 0.0131 | 0.8386 |
|  | DS5 | 0.0325 | 0.0048 | 0.0693 | NaN | 0.0125 | 0.8504 |

***Summary for RQ1:*** *There is a non-linear relationship between Coverageability and source code metrics. The learning efficiency of non-linear regression models is suitable enough to predict Coverageability in practice. The best machine learning model is VoR, the combination of MLPR, RFR, and HGBR, with an MAE of 0.032, MSE of 0.004, and $R^2$-score of 0.855.*

In addition to Table 6, Figure 6 compares the MSE, given in Equation 9, and $R^2$-score, given in Equation 13, of the four non-linear regression models on different datasets. According to the datasets discussed in Table 4, the following results are observed:

- Automatic feature selection decreases the regression models' performance, *e.g.*, $R^2$-score, and subsequently increases the model error, *e.g.*, MSE. It confirms that the most considered source code metrics contain helpful information about Coverageability.
- The package-level metrics harm the regression models' performance. However, their impact is negligible. For the best prediction model, *i.e.*, VoR, the result of DS3 and DS1 are similar. It allows us to eliminate package-level metrics and speed up the training process without losing performance.
- File-level lexical metrics and also sub-metrics have positive impacts on estimating Coverageability. However, their impact is not significant. For the best model, VoR, the difference in $R^2$-score between DS3 and DS4 is about 0.016, and between DS3 and DS5 is about 0.004. Therefore, our proposed metrics can be used to build a more accurate model to predict Coverageability.

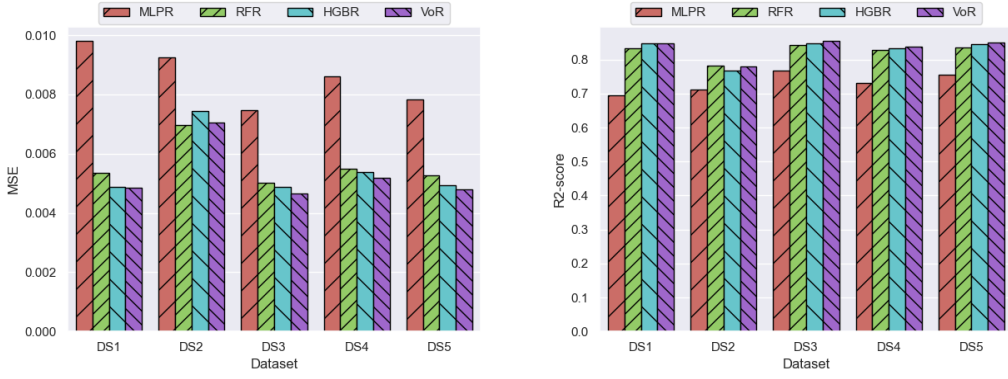

Figure 6. Comparison of different datasets on Coverageability prediction

> **Summary for RQ2:** *The proposed sub-metrics and file-level lexical metrics positively affect the performance of learning class Coverageability, while package-level metrics have negative impacts. Automatic feature selection should not be used when building the Coverageability models since it increases the prediction error rate.*

### 4.4 Coverageability analysis

We discuss the distribution of class Coverageability on many Java projects and compare Coverageability and code coverage prediction models' performance.

#### 4.4.1 Coverageability distribution

Figure 7 shows the distribution of Coverageability values obtained for the classes in SF110. The vertical axis indicates the logarithm of the number of classes with a specific Coverageability value, denoted by the horizontal axis, on a logarithmic scale. Coverageability is computed by Equation 4 using $b = 1$. MeanCoverage, computed by Equation 3, indicates Coverageability when $b = |\tau(X, C)|$. In other words, Figure 7 illustrates the frequency of classes between the upper and lower boundaries of the Coverageability determined by the test budget. The following results are observed:

- Most Java classes in our dataset achieve a high statement and branch coverage; therefore, their mathematical expected coverage is complete or close to 1, while their Coverageability with a minimum number of test cases, *i.e.*, $b = 1$, is below 30%. It means that they are complex classes and require many test cases to be covered completely.
- Coverageability captures both the effectiveness and cost of testing within an interpretable interval of (0,1). It considers the cost of generating and executing each test case.

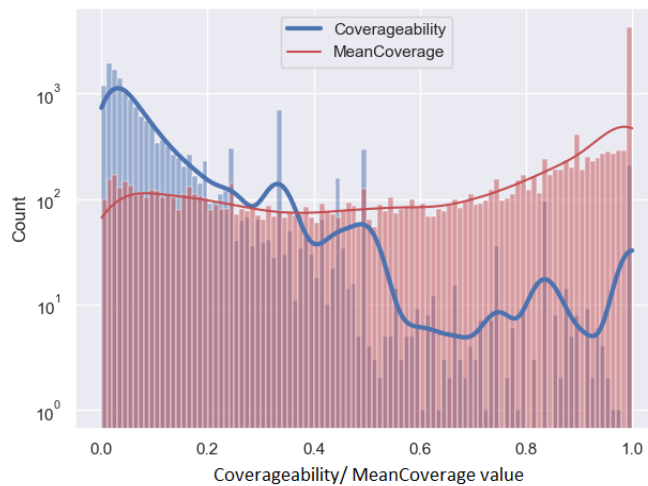

Figure 7. Distribution of Coverageability between zero and 1

#### 4.4.2 Coverageability versus code coverage

Figure 8 compares the mean squared error of Coverageability, statement coverage, branch coverage, and MeanCoverage prediction made by different regression models on different datasets. The $R^2$-score of the models is shown in Figure 9.

The following results are observed:

- The MSEs of our Coverageability prediction models are less than those of the statement, branch, and MeanCoverage. Machine learning algorithms can learn Coverageability more effectively than learning branch, statement, and the average of statement and branch coverage. Coverageability takes into account the cost of testing as an uncertainty factor found in the test suites. The size of the test suite makes Coverageability a suitable measure to be predicted by source code metrics.
- The MSE of the regression models to predict MeanCoverage, given in Equation 3, is less than the same models trained to predict statement and branch coverage. Therefore, the arithmetic mean of statement and branch coverage can be regarded as a relatively better criterion for determining test effectiveness. Two classes with almost the same source code metrics may achieve different statement and branch coverage. The mathematical expectation of coverage solves the imbalanced distribution of statements in branches and allows machine learning to train effectively on such programming structures.
- The $R^2$-score of the Coverageability prediction models is more than the test criteria prediction models, making these models suitable to be used in practice to predict the cost and effectiveness of testing.

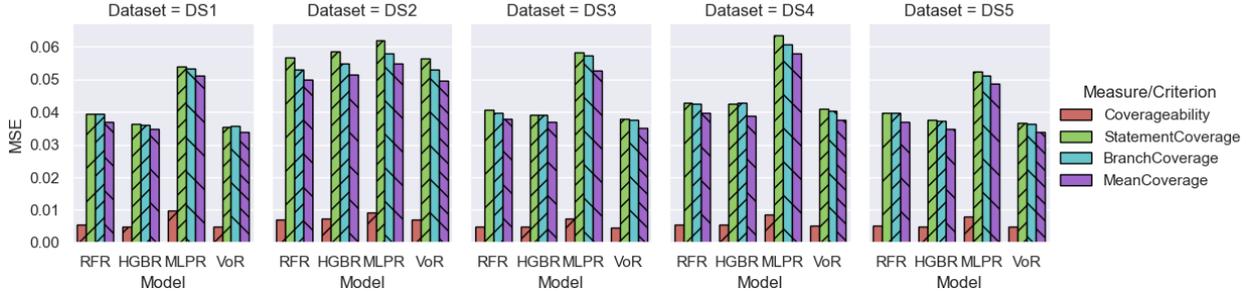

Figure 8. Mean squared error for Coverageability and code coverage prediction.

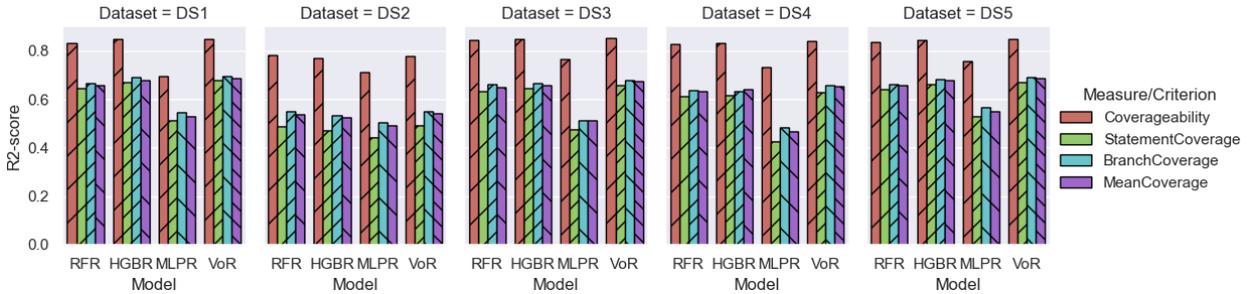

Figure 9. $R^2$-score for Coverageability and code coverage prediction.

### 4.4.3 Code coverage prediction

Table 7 compares the evaluation metrics of the prediction models for statement and branch coverage with the results reported by Grano et al. [7] for branch coverage. Grano et al. [7] have trained and evaluated their machine learning models on 7 Java projects with 3105 classes. They have also extracted 79 source code metrics to build the prediction models and reported RFR as the best machine learning regression model. We found that the best model is an ensemble of different regression models. In summary, compared to Grano et al. approach:

- MAE and MSE of our models are lower by at most 0.0578 and 0.0284, respectively.
- The $R^2$-score of our model is higher by 0.2071.
- In total, our models outperform the previous code coverage prediction models.
- Extending the number of projects and, subsequently, Java classes as dataset samples and increasing the number of source code metrics as dataset features improve the model performance.

Table 7. Comparison of the performance of code coverage prediction models

| Criterion | MAE | MSE | RMSE | MSLgE | MdAE | $R^2$-score |
|---|---|---|---|---|---|---|
| Statement Coverage | 0.1362 | 0.0378 | 0.1944 | NaN | 0.0910 | 0.6595 |
| Branch Coverage | 0.1332 | 0.0376 | 0.1939 | 0.0170 | 0.0848 | 0.6801 |
| Branch Coverage [7] | 0.1910 | 0.0660 | — | 0.0300 | 0.1520 | 0.4730 |

> **Summary for RQ3:** *Coverageability prediction models outperform individual test criteria prediction models in terms of MAE, MSE, and $R^2$-score metrics. Source code metrics could best determine the Coverageability in comparison with predicting statement and branch coverage. The uncertainty of test suites must be taken into account when predicting the cost and effectiveness of tests.*

## 4.5 Model inspection and important metrics

In our final experiment, we study the impact of each source code metric on Coverageability to determine the most influential metrics affecting class Coverageability. One of the Coverageability model's primary design goals is to determine the vital source code metrics affecting test effectiveness. Previous researches address this problem with very naïve heuristics, such as manually selecting or defining a set of metrics to measure testability [39], [51]. In fact, they manually select a subset of source code metrics that they guess are appropriate for determining test effectiveness. It is a very rigorous method to find trustable testability metrics manually.

We act differently and allow all available metrics to participate in the process of the Coverageability prediction. Using model inspection techniques, we measure the contribution of each metric to Coverageability prediction. Thereby, we can automatically identify source code metrics that highly affect the Coverageability of the class under test. To this aim, we applied permutation importance [4] to one of our prediction models VoR using the DS3 dataset. We chose the VoR model because it was relatively more accurate than the other four prediction models we built. We repeated the permutation process 50 times and recorded the test set's $R^2$-score (accuracy) changes. Afterward, we sorted the metrics according to their impact on the prediction results.

Figure 10 illustrates the box-plot of the top 15 most influential metrics on the Coverageability prediction. The boxes' data points show the exact value of changes in the $R^2$-score after shuffling a specific feature. Figure 11 depicts important metrics as the web of internal quality attributes affecting Coverageability. We constructed the plot by determining the quality subject of each important metric in Figure 10 and counted the metrics for each attribute. The following results are observed:

- The most important metric affecting Coverageability is Class Strict Cyclomatic Complexity (CSCCS), computed by summing up the Strict Cyclomatic Complexity of all methods in a class. The Strict Cyclomatic Complexity enumerates logical conjunction in conditional expressions and adds 1 to the complexity for each of their occurrences. For instance, the Cyclomatic Complexity of the statement `if (x && y || z)` is 1 while Strict Cyclomatic Complexity is 3.
- Eight and three of the fifteen metrics are relevant to the size and complexity, respectively. It implies that size and complexity attributes affect Coverageability more than the other metrics.
- Cohesion and coupling metrics do not belong to the top 15 important metrics. However, we observed that metrics such as FANOUT are amongst the top 20 influential metrics.
- Model inspection techniques could also be applied to determine each software metrics' impact on any test adequacy criterion.

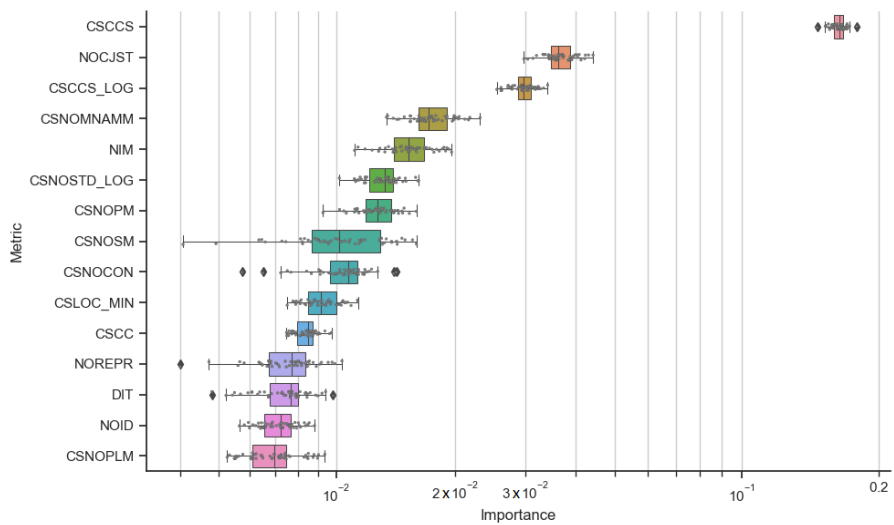

Figure 10. Top 15 most influential source metrics affecting Coverageability

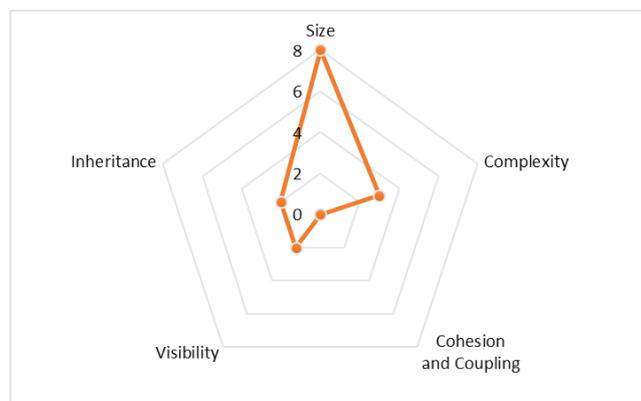

Figure 11. Quality attributes corresponding to important Coverageability metrics

To determine the impact of each influential metric on Coverageability, we computed the Pearson correlation coefficient between the class Coverageability and each metric. A positive correlation means that an increase in the value of a metric increases the Coverageability, while a negative Correlation means that increasing the value of the metric worsens Coverageability. We also reported the *P*-value to demonstrate the meaningfulness of our inference. If *P*-value for a given correlation coefficient is greater than .05, the correlation is not statistically significant. In this respect, we decided to report the Unknown relationship between each of these source code metrics and Coverageability. Table 8 shows the correlation coefficient and its P-value between each of the 15 selected metrics and Coverageability. We observe the following results:

- Almost all source code metrics positively impact Coverageability, which means that as the metric improves, the Coverageability also improves. All these metrics improve as their value decreases, while an improvement in Coverageability implies an increase in its value. That is why the correlation between Coverageability and each of these metrics, as shown in Table 8, is negative.
- The only metric that positively correlates with Coverageability is the Depth of the Inheritance Tree (DIT). It means that as the DIT of a class increases, its Coverageability increases. *A class with a deep inheritance tree presumably delegates its responsibility to its ancestors, due to which Coverageability improves*.
- There is no statistically significant correlation between Coverageability and the three metrics of Number Of Inherited Methods (NIM), Class Number Of Constructors (CSNOCON), and Class Minimum Line Of Code (CSLOC_MIN) since their P-values are greater than 0.05. It means that more samples are required to identify the correlations between these metrics and Coverageability.

Table 8. Correlation between Coverageability and most influential source code metrics

| Metric | Correlation | *P*-value | Impact on Coverageability |
| --- | --- | --- | --- |
| CSCCS | -0.31905 | << .001 | Negative |
| NOCJST | -0.26063 | < .001 | Negative |
| CSCCS_LOG | -0.67403 | << .001 | Negative |
| CSNOMNAMM | -0.30092 | < .001 | Negative |
| NIM | -0.01453 | .064 | Unknown |
| CSNOSTD_LOG | -0.66080 | < .001 | Negative |
| CSNOPM | -0.14407 | < .001 | Negative |
| CSNOSM | -0.12431 | < .001 | Negative |
| CSNOCON | -0.00820 | .297 | Unknown |
| CSLOC_MIN | +0.07374 | < .071 | Unknown |
| CSCC | -0.31872 | < .001 | Negative |
| NOREPR | -0.30070 | < .001 | Negative |
| DIT | +0.14445 | < .001 | Positive |
| NOID | -0.33038 | < .001 | Negative |
| CSNOPLM | -0.21556 | < .001 | Negative |

**Summary for RQ4:** *Source code metrics concerning the size and complexity are the most influential metrics affecting Coverageability. These metrics negatively impact Coverageability. Depth of Inheritance Tree (DIT) is the only metric that positively affects Coverageability by a low correlation coefficient of +0.15. We could not determine the impacts of the NIM, CSNOCON, and CSLOC_MIN metrics on Coverageability since relatively more data samples were required.*

# 5 Threats to validity

Certain factors may affect the construction, internal and external validity of our proposed Coverageability prediction model. Firstly, we apply source code metrics and coverage metrics to construct our prediction model. Statement and branch coverage are among the most commonly used metrics to evaluate test effectiveness in practice [33]. However, other coverage metrics such as mutation score and number of assertions may provide a more precise assessment of test effectiveness.

Secondly, the most critical threats to the internal validity of our prediction model are the stochastic nature of both evolutionary algorithms used for test data generation and machine learning algorithms. We repeated the test suite generation process five times for each project using default parameters with different random seeds to ensure the reliability of the EvoSuite. We then compute the average of results obtained for each criterion during each iteration. There is still an opportunity to run EvoSuite with different hyperparameters and create a more reliable test. We use a grid search strategy with five-fold cross-validation to avoid overfitting and find the best possible regression models. Other machine learning models and parameters should also experiment with to maximize the learning gain.

Finally, the main threat to the external validity of our Coverageability model is related to extending our approach to other software and programming languages (PLs). We conducted our experiments on SF110 [6], which is big enough to generalize all types of software. SF110 contains 110 Java projects and 23,000 classes. We also used a set of 296 source code metrics to build our prediction models. Our models can be used to predict the Coverageability of programs written in other programming languages, especially those similar to Java, *i.e.*, C++ and C#. Nevertheless, it is required to evaluate the performance of learned models in different programming languages to ensure their applicability in practice.

# 6 Conclusion

Coverageability is a new concept introduced in this paper. Coverageability as an inherent feature of the software is the probability of a class being tested effectively considering the test budget. Coverageability is computable in terms of coverage provided by the test suite, with an acceptable test budget. Regressor models can best predict Coverageability by mapping a class represented as a vector of metrics to its expected test coverage. The top 5 most influential source code metrics affecting Coverageability are Class Cyclomatic Complexity (CSCC), Number Of Conditional Jump Statement (NOCJST), Number Of Not Accessor and Mutator Methods (CSNOMNAMM), Number of Inherited Methods (NIM), and Number Of Statements (CSNOST) which belong to size, complexity, and inherent quality attributes. It is shown in this paper that an ensemble of four regression models can predict Coverageability accurately. The prediction accuracy in the $R^2$-score is about 85% which is relatively high compared to the models that solely predict code coverage. Our experimental results show that coverage metrics may complement each other to provide a more suitable assessment of test effectiveness. The arithmetic average of branch and statement coverage provides a relatively more accurate estimation of Coverageability than each of the coverage metrics on their own. In future work, we plan to improve Coverageability automatically by applying appropriate code refactorings to improve software metrics affecting Coverageability. We are planning to use search-based refactoring techniques to maximize Coverageability. To this aim, the Coverageability prediction model will be used as an objective for the search-based algorithm.

**CONFLICT OF INTEREST**

The authors declare that they have no conflict of interest.